\documentclass[12pt,reqno]{article}
\usepackage{amsmath,amssymb,amsthm}
\usepackage{CJK}
\usepackage{amsfonts,amssymb}
\usepackage{cite}
\usepackage{longtable}
\usepackage{rotating}
\usepackage{multirow}
\usepackage{setspace}
\usepackage{float}
\usepackage{threeparttable}
\usepackage{graphicx}
\usepackage{indentfirst}
\usepackage{slashbox,pict2e}
\usepackage{multirow}
\usepackage{lineno}
\usepackage{booktabs}
\usepackage[titletoc]{appendix}
\usepackage{mathrsfs}
\usepackage[pdfstartview=FitH,%
bookmarksnumbered=true,bookmarksopen=true,%
colorlinks=true,pdfborder=001,citecolor=blue,%
linkcolor=red,urlcolor=blue]{hyperref}
\usepackage[numbers,sort&compress]{natbib}
\usepackage{booktabs}
\usepackage{psfrag}
\usepackage{amsmath}
\usepackage{graphicx}
\usepackage{subfigure}
\usepackage[titletoc]{appendix}
\newtheorem{theorem}{Theorem}

\begin{document}

\title
{Modeling and Computation of Mean Field Equilibria in Producers' Game with Emission Permits Trading}
\author{Shuhua Chang$^{1,2}$,   Xinyu Wang$^{1}$,  \mbox{and} Alexander Shananin$^3$\\
\\
$^1$  Research Center for Mathematics and Economics \\
 Tianjin University of Finance and Economics \\
 Tianjin 300222, China\\
 $^2$ Institute of Policy and Management\\
 Chinese Academy of Sciences, Beijing 100190, China\\
 $^3$ Department of Control and Applied Mathematics\\
 Moscow Institute of Physics and Technology, Russia\\
\texttt{Corresponds to shuhua55@126.com
}\thanks{This project was supported in part by the National Basic Research Program (2012CB955804), the Major Research Plan of the National Natural Science Foundation of China (91430108), the National Natural Science Foundation of China (11171251), the Russian Foundation for Basic Research (14-07-00075), and the Major Program of Tianjin University of Finance and Economics (ZD1302).}
}
\date{\today}
\maketitle
\begin{abstract}
In this paper, we present a mean field game to model the production behaviors of a very large number of producers, whose carbon emissions are regulated by government. Especially, an emission permits trading scheme is considered in our model, in which each enterprise can trade its own permits flexibly. By means of the mean field equilibrium, we obtain a Hamilton-Jacobi-Bellman (HJB) equation coupled with a Kolmogorov equation, which are satisfied by the adjoint state and the density of producers (agents), respectively. Then, we propose a so-called fitted finite volume method to solve the HJB equation and the Kolmogorov equation. The efficiency and the usefulness of this method are illustrated by the numerical experiments. Under different conditions, the equilibrium states as well as the effects of the emission permits price are examined, which demonstrates that the emission permits trading scheme influences the producers' behaviors, that is, more populations would like to choose a lower rather than a higher emission level when the emission permits are expensive.
\end{abstract}

\noindent
{\bf Keywords.} Mean field games, Producers' behaviors, Emission permits trading, Fitted finite volume method, Equilibrium states.

\section{Introduction}
Due to the rapid development of economy and society, every one should analyse and get a clear understanding of a very complex system before making a decision. In particle physics, mean field theory can be regarded as a useful and effective tool to study the performance of large and complex stochastic model. It focuses mainly on a single particle and assumes that the interactions of this particle with its neighboring particles is determined by the mean field, in which the inter-particle interactions must be sufficiently ``weak'' or ``regular'' and each particle tends to be infinitesimal, i.e., the number of particles tends to infinity.

Based on the above features of mean field theory, Lasry and Lions first defined and developed mean field game in their three papers \cite{LL06a, LL06b, LL07}. Different from the standard game theory in which the number of players is finite, {\color {red}such as \cite{ZJZ14, WWP14, SWTF13, WSP13}}, mean field game studies the limit case of a game with $N$ players as $N$ goes to infinity, which implies a continuum of agents. Owing to this, mean field game can be used to deal with the problems, which should be summed up by an untractable system of HJB equations in general differential games with $N$ players, where $N$ is very large.

Several researchers have studied the theory and applications of mean field game in recent years \cite{HT01,G09a, G09b, CPTD12, RRST07}. Especially, some interesting and meaningful works relating the mean field game to economics have been done. Gomes et al. present effective numerical methods for two-state mean field games and discuss a number of illustrative examples in socio-economic sciences in \cite{GVW14}. Besides, in \cite{LST10} Lachapelle et al. consider a technology choice problem with externalities and  economy of scale by using a mean field game stylized model. They introduce a monotonic algorithm to find the mean field equilibria and describe the multiplicity of equilibria.

However, to our best knowledge, there are very few studies on mean field games to take emission permits trading into consideration. For the past few years, the issues on climate change and emission reduction have attracted the attention of politicians and scholars from all over the world. In order to mitigate climate change and improve global environment, some emission permits trading markets have emerged and are developing prosperously. At the same time, a large number of articles have studied the emission permits price theoretically and empirically \cite{CWP13, DPM09, HU13, SUW08, CW15}. Moreover, some differential game models about production decision also include the emission permits trading scheme \cite{L14, BHV08}.

Motivated by the those mentioned above, in this paper, we build a mean field game model, in which the revenues from emission permits trading are included, to study the productive behaviors of a continuum of agents under the background of climate change and emission reduction. In our model, each agent is anonymous and the interactions among agents are mean field type. They can obtain an initial quota from the emission permits trading scheme, and purchase the emission permits from the market compulsively if the quota is insufficient or sell the unused emission permits to others. The equilibrium of mean field game can be reached by solving two coupled partial differential equations, one of which is a backward HJB equation satisfied by the adjoint state, and the other one is a forward Kolmogorov equation satisfied by the density of agents.

Some discussions about the numerical algorithms of the above coupled system have been made for the past few years \cite{GVW14, LST10, AD10, G12a, G12b}. All of these methods are based on a finite difference scheme. In this paper, we propose a so-called fitted finite volume method to solve the coupled equations model established by ourselves. The innovation of this method is in that it couples a finite volume formulation with a fitted local approximation to the solution. On the one hand, we implement the local approximation through solving a sequence of two-point boundary value problems defined on each element. On the other hand, the finite volume method possesses a special feature of the local conservativity of the numerical flux. The main advantage of this discrete method is that the system matrix of the resulted discrete equation is an $M$-matrix, which guarantees that the discretization is monotonic and the discrete maximum principle is satisfied. The finite volume method, except for the fundamental fluid dynamics problem in which it performs very well, has been used in many fields and is becoming more and more popular. See, for instance, \citep{Wang} for degenerate parabolic problems, \citep{Leveque} for hyperbolic problems, and \citep{Liu} for elliptic problems.

The paper is organized as follows. In Section $2$, a mean field game model is established, and the coupled partial differential equations are presented. Then, a so-called fitted finite volume method is proposed for the discretization of the equations in Section $3$. In Section $4$, some numerical experiments are performed to illustrate the efficiency and usefulness of the numerical method, and the effects of the parameters on the density of population are also showed in this section. Finally, concluding remarks are given in Section $5$.

\section{The mean field game model}
\subsection{The states of agents}
For the purpose of illustrating the interactions among the players in a commitment period, we propose a finite-horizon mean field game framework. We focus on a very large economy and consider a continuum of agents. Each agent is a producer whose carbon emission is regulated by governments. In our settings, all the producers have the same capacity in production. They are anonymous but different in their initial production states which follow a given initial probability density function. In addition, the interactions among the agents are mean field type. That is to say, a given agent cannot influence the distribution of all players' states and therefore the decisions of others by itself. However, it can produce an effect on the information which is used by others to make decisions. These ideas, as well as the name of $mean$  $field$, come from particle physics, and the particles are replaced by rational agents here.

Let $Q(t)$ denote the production of each agent during the period of $[0,T],$ where $T$ is the maturity of the game. This production leads to an amount of by-products, namely emissions $E(t)$. Generally speaking, an increase in production can result in more emissions and vice versa. So, it is reasonable to use the emission as a state variable instead of production in this paper. In fact, the emission level can be also treated as a product portfolio. The dynamics of the agent's emission is given by the following controlled process:
\begin{equation}\label{emission}
dE_{t}=-\tau(t,E) dt +\sigma dW +dN_{t}(E_{t}),
\end{equation}
where $\tau(t,E)$ is a control variable, and can be interpreted as the level of emission reduction. The term $\sigma dW$ stands for the stochastic disturbance of emission resulting from the technological innovations, market fluctuations, and some other uncertain factors. The constant $\sigma$ is a noise parameter and denotes the volatility of emission, and $dW$ is the increment to the standard Brownian process. In addition, the emission $E_t$ is restricted in $[E_{\min}, E_{\max}]$ by the reflection part $dN_{t}(E_{t}),$ in which $N_t$ is a monotonic continuous nondecrease function. For more details about this formulation, please see \cite{LST10} and \cite{F85}.

Taking advantage of the dynamics of emission \eqref{emission}, we can obtain the forward Kolmogorov equation satisfied by the density of agents $m(t,E)$ for $t\in[0,T]$ and $E\in[E_{\min},E_{\max}]$:
\begin{equation}\label{equation_m}
\begin{cases}
\displaystyle\frac{\partial m}{\partial t}-\displaystyle\frac{1}{2} \sigma^2 \displaystyle\frac{\partial^2 m}{\partial E^2}+\frac{\partial}{\partial E}(-\tau(t,E) m)=0, ~~ t\in(0,T] ~~\mbox{and} ~~ E\in(E_{\min},E_{\max}), \\[2mm]
m(0,E)=m_0, ~~ E\in[E_{\min},E_{\max}],\\[2mm]
m'(t,E_{\min})=m'(t,E_{\max})=0, ~~ t\in[0,T],
\end{cases}
\end{equation}
where $m_0$ is the initial density. This equation is also called the Fokker-Planck equation in physical literature. The detailed derivation of this equation can be found in \cite{LST10}, \cite{KP92}, and \cite{Smears}.

\subsection{The revenues of agents}
Every agent in our model should choose the best emission reduction level $\tau$ to maximize its own net revenues, which consists of three parts, namely, the production revenue, the cost of emission reduction, and the net revenue from the emission permits trading scheme.

Firstly, each player's revenue arising from the production can be represented by an increasing concave function $R(Q(t))$. Following \cite{L14} and \cite{Breton}, we assume that the relationship between production and emissions is linear, and the production revenue function can be expressed by the following quadratic functional form in terms of emissions:
\begin{equation*}
R(E)=\frac{AE-\frac{1}{2}E^2}{c_1+c_2m(t,E)},
\end{equation*}
where $A=E_{\max}$, $c_1,$ and $c_2$ are constants. Under this assumption, the marginal revenues are positive and decreasing. In addition, the production revenue is decreasing with respect to the density of agents $m$, which can be explained by the economical concept ``negative externality'', that is to say, an agent should face more intense competition and receive fewer revenues if it chooses the same state as others' ones.

Secondly, the cost of emission reduction should be
\begin{equation*}
C(\tau_t)=\frac{\tau_t^2}{2}.
\end{equation*}
This quadratic form guarantees increasing marginal mitigation cost.

Finally, the gains from emission permits trading are expressed by
\begin{equation*}
T(E)=S(t)*(E-E_0),
\end{equation*}
where $S(t)$ denotes the emission permits price and $E_0$ is the initial quota. The trading volume of emission permits $E-E_0>0$ means that an agent purchases the emission permits from the market, and $E-E_0<0$ means that an agent sells the unused emission permits to others, respectively.
\subsection{The maximization problem and the optimal conditions}
Now, we are in the position to define our maximization problem for each agent. That is, the objective functional and the constraint condition are as follows:
\begin{equation}\label{initial_op}
\begin{aligned}
&\max_{\tau_t}\mathbb{E}\Bigg\{\int_0^{T}e^{-rt}\left[\frac{AE-\frac{1}{2}E^2}{c_1+c_2m(t,E)}-\frac{\tau_t^2}{2}
-S(t)(E-E_0)\right]dt\Bigg\},\\
&\mbox{subject to} \quad
dE_t=-\tau_t dt +\sigma dW +dN_t(E_t), ~~  E(0)=E_0,
\end{aligned}
\end{equation}
where $r$ is the risk-free discount rate, and $t=0$ is the initial time.

According to \eqref{equation_m}, this problem can be reformulated as follows:
\begin{equation}\label{derived_op}
\left\{
\begin{aligned}
&\max_{\tau_t}\int_0^{T}e^{-rt}\bigg\{\int_{E_{\min}}^{E_{\max}}\left(\frac{AE-\frac{1}{2}E^2}{c_1+c_2m(t,E)}-\frac{\tau_t^2}{2}
-S(t)(E-E_0)\right)m(t,E)dE\bigg\}dt,\\
&\frac{\partial m}{\partial t}-\frac{1}{2} \sigma^2 \frac{\partial^2 m}{\partial E^2}+\frac{\partial}{\partial E}(-\tau(t,E) m)=0, ~~ m'(t,E_{\min})=m'(t,E_{\max})=0,
\end{aligned}
\right.
\end{equation}
with the initial condition $m(0,E)=m_0$.

The solution of the above problem should correspond to the equilibria of mean field game, in which every producer is atomized and has rational expectations.

Next, we will show the process of obtaining the optimal conditions for problem \eqref{derived_op}. To begin with, we multiply $v$ on both sides of equation \eqref{equation_m} and integrate by parts to obtain the following weak form:
\begin{align*}
\int_{E_{\min}}^{E_{\max}}\left(v(T,E)m(T,E)-v(0,E)m(0,E)\right)dE
=\int_0^T\int_{E_{\min}}^{E_{\max}}\left(\frac{\partial v}{\partial t}+\frac{1}{2}\sigma^2\frac{\partial^2 v}{\partial E^2}-\tau\frac{\partial v}{\partial E}\right)mdEdt
\end{align*}
for every $v\in C_c^{\infty}([E_{\min},E_{\max}]\times [0,T])$, where
$$
C^{\infty}_c(\Omega)=\left\{u\in C^{\infty}(\overline{\Omega});\,\mbox{supp}\,{u}=\overline{\{x;\,u(x)\neq 0\}}\subset\Omega\right\}.
$$
Then, the Lagrangian of \eqref{derived_op} should be
\begin{align*}
L(m,\tau,v)&=\int_0^{T}e^{-rt}\bigg\{\int_{E_{\min}}^{E_{\max}}\left(\frac{AE-\frac{1}{2}E^2}{c_1+c_2m(t,E)}-\frac{\tau_t^2}{2}
-S(t)(E-E_0)\right)m(t,E)dE\bigg\}dt\\
&+\int_0^T\int_{E_{\min}}^{E_{\max}}\left(\frac{\partial v}{\partial t}+\frac{1}{2}\sigma^2\frac{\partial^2 v}{\partial E^2}-\tau\frac{\partial v}{\partial E}\right)mdEdt\\
&-\int_{E_{\min}}^{E_{\max}}\left(v(T,E)m(T,E)-v(0,E)m(0,E)\right)dE.
\end{align*}
Taking the derivatives of Lagrangian with respect to $m$, $\tau$ and $v$, respectively, we can obtain the following system:
\begin{equation}\label{equilibira_system}
\left\{
\begin{aligned}
&\frac{\partial v}{\partial t}+\frac{1}{2} \sigma^2 \frac{\partial^2 v}{\partial E^2}-\tau \frac{\partial v}{\partial E}-rv+f(E,\tau,m,t)=0, ~~ v(T,E)=0,\\
&\tau=-\frac{\partial v}{\partial E},\\
&\frac{\partial m}{\partial t}-\frac{1}{2} \sigma^2 \frac{\partial^2 m}{\partial E^2}+\frac{\partial}{\partial E}\left(-\tau m\right)=0, ~~ m(0,E)=m_0, ~~ m'(t,{E_{\min}})=m'(t,E_{\max})=0,
\end{aligned}
\right.
\end{equation}
where $f(E,\tau,m,t)=-\frac{\tau^2}{2}-S(E-E_0)+\frac{AE-\frac12E2}{c_1+c_2m}-\frac{c_2m(AE-\frac12E^2)}{(c_1+c_2m)^2}$.
Clearly, we can see that this system consists of two coupled equations, one of which is the backward HJB equation and the other one is the forward Kolmogorov equation. In fact, the solution of this system is the mean field equilibrium of our game.
\section{Numerical methods}
In this section, we will present a numerical method to discretize the first equation of \eqref{equilibira_system} for the reason that it is difficult to solve the equation analytically. In fact, here a fitted finite volume method will be employed. Also, it will be shown that the system matrix of the resulting discrete equations is an $M$-matrix, which guarantees that the discretization is monotonic and the discrete maximum principle is satisfied, such that the scheme has a unique solution. Besides, a two-level implicit time-stepping method is used to implement the time-discretization. Since the structure of Kolmogorov equation is similar to the HJB equation, here we only discuss the latter to save the space.
\subsection{The fitted finite volume method for spatial discretization}
A defined mesh for $(E_{\min},E_{\max})$ is significant in the process of discretization. We first divide the intervals $I=(E_{\min},E_{\max})$ into $N$ sub-intervals:
$$
I_{i}:=(E_i,E_{i+1}), ~~i=0,1,\cdots,N-1,,
$$
in which
$$
E_{\min}=E_0<E_1<\cdots<E_{N}=E_{\max}.
$$
For each $i=0,1,\cdots,N-1$, we define another partition of $I$ by letting
$$
E_{i-\frac{1}{2}}=\frac{E_{i-1}+E_{i}}{2},\;
E_{i+\frac{1}{2}}=\frac{E_{i}+E_{i+1}}{2}.
$$
To keep completeness, we also define $E_{-\frac{1}{2}}=E_{\min}$ and $E_{N+\frac{1}{2}}=E_{\max}$. The step is defined by $h=E_{i+\frac{1}{2}}-E_{i-\frac{1}{2}}$ for each $i=0,1,\cdots,N$.

Then, for the purpose of formulating finite volume scheme, we write the first equation of \eqref{equilibira_system} in the following divergence form:
\begin{equation}\label{div}
-\frac{\partial v}{\partial t}-\frac{\partial}{\partial E}\left(a\frac{\partial v}{\partial E}+bv\right)+cv-f(E,\tau,m,t)=0,
\end{equation}
where
\begin{align*}
&a=\frac12\sigma^2,\quad b=-\tau,\\
&c=r+\frac{\partial b}{\partial E}=r-\frac{\partial \tau}{\partial E}.
\end{align*}

It follows from integrating equation \eqref{div} over $(E_{i-\frac12},E_{i+\frac12})$ and applying the mid-point quadrature rule to the resulting equation that
\begin{eqnarray}\label{mid-qua}
-\frac{\partial v_i}{\partial t}l_i-\left[\rho(v)|_{E_{i+\frac12}}-\rho(v)|_{E_{i-\frac12}}\right]+c_iv_il_i-f_il_i=0
\end{eqnarray}
for $i=1,2,\cdots,N-1$, where $l_i=E_{i+\frac{1}{2}}-E_{i-\frac{1}{2}}$, $c_{i}=c(E_i,t)$, $v_{i}=v(E_i,t)$, $f_{i}=f(E_i,\tau,m,t),$ and $\rho(v)$ is a flux associated with $v$ defined by
\begin{equation}\label{flux}
\rho(v)=a\frac{\partial v}{\partial E}+bv.
\end{equation}
Now, we focus on deriving an approximation to the flux at mid-point, $E_{i+\frac12}$, of the interval $I_i$ for all $i=0,1,\cdots,N-1$. Consider the following two-point boundary value problem:
\begin{subequations}
\begin{gather}\label{ode_a}
(av'+b_{i+\frac12}v)'=0, ~~ E\in I_i,\\\label{ode_b}
v(E_i)=v_i, ~~ v(E_{i+1})=v_{i+1},
\end{gather}
\end{subequations}
where $b_{i+\frac12}=b(E_{i+\frac12},t)$. A first-order ordinary differential equation can be obtained by integrating both sides of equation \eqref{ode_a}:
$$
\rho_i(v)=av'+b_{i+\frac12}v=C_1,
$$
where $C_1$ is an arbitrary constant and can be determined by solving the above constant coefficient two-point boundary problem analytically as follows:
\begin{equation}\label{rho}
\rho_i(v)=C_1=b_{i+\frac12}\frac{e^{\alpha_iE_{i+1}}v_{i+1}-e^{\alpha_iE_{i}}v_{i}}{e^{\alpha_iE_{i+1}}-e^{\alpha_iE_{i}}},
\end{equation}
where $\alpha_i=b_{i+\frac12}/a$.

By using \eqref{rho}, we can define a piecewise constant approximation to $\rho(v)$ by $\rho_h(v)$ satisfying
\begin{equation}\label{rho1}
\rho_h(v)=\rho_i(v) ~~ \mbox{if} ~~ x\in I_i
\end{equation}
for $i=0,1,\cdots,N-1$.

Thus, substituting \eqref{rho1} into \eqref{mid-qua}, we obtain the following results:
\begin{equation}\label{dis_form}
-\frac{\partial v_i}{\partial t}l_i+e_{i,i-1}v_{i-1}+e_{i,i}v_{i}+e_{i,i+1}v_{i+1}=f_il_i,
\end{equation}
where
\begin{align}\label{eleft}
e_{i,i-1}&=-b_{i-\frac12}\frac{e^{\alpha_{i-1}E_{i-1}}}{e^{\alpha_{i-1}E_{i}}-e^{\alpha_{i-1}E_{i-1}}},\\\label{ecenter}
e_{i,i}&=b_{i-\frac12}\frac{e^{\alpha_{i-1}E_{i}}}{e^{\alpha_{i-1}E_{i}}-e^{\alpha_{i-1}E_{i-1}}}+b_{i+\frac12}\frac{e^{\alpha_{i}E_{i}}}{e^{\alpha_{i}E_{i+1}}-e^{\alpha_{i}E_{i}}}+c_il_i,\\\label{eright}
e_{i,i+1}&=-b_{i+\frac12}\frac{e^{\alpha_{i}E_{i+1}}}{e^{\alpha_{i}E_{i+1}}-e^{\alpha_{i}E_{i}}}.
\end{align}
\subsection{The implicit difference method for time discretization}
Next we embark on the time-discretization of the system \eqref{dis_form}. To this purpose, we first rewrite equation \eqref{dis_form} as
\begin{equation}\label{dis_sys}
-\frac{\partial v_{i}}{\partial t}l_i+D_iv=f_{i}l_{i},
\end{equation}
where
\begin{align*}
D_1&=(e_{1,1},e_{1,2},0,\cdots,0),\\
D_{i}&=(0,\cdots,0,e_{i,i-1},e_{i,i},e_{i,i+1},0,\cdots,0), ~~ i=2,3,\cdots,N-2,\\
D_{N-1}&=(0,\cdots,0,e_{N-1,N-1},e_{N-1,N}).
\end{align*}
We select $K-1$ points numbered from $t_1$ to $t_{K-1}$ between $0$ and $T,$ and let $T=t_0$, $t_K=0$ to form a partition of time $T=t_0>t_1>\cdots>t_K=0$. Then, the full discrete form of equation \eqref{dis_sys} can be obtained by applying the two-level implicit time-stepping method with a splitting parameter $\theta \in [\frac{1}{2},1]$ to it:
\begin{equation}\label{dis_sys2v}
\begin{split}
(\theta D(E,\tau,t^{k+1}) + G^{k})v^{k+1}&=\theta f(E,\tau,t^{k+1})l_i+(1-\theta)f(E,\tau,t^{k})l_i\\
&+(G^k-(1-\theta)D(E,\tau,t^{k}))v^k,
\end{split}
\end{equation}
where
\begin{align*}\label{dis_F}
D&=(D_1,D_2,\cdots,D_{N-1})^{\top},\\
G^k&=\mbox{diag}\,(-l_1/\Delta t_k,\cdots,-l_{N}/\Delta t_k)^{\top},
\end{align*}
for $k=0,1,\cdots,K-1$. Note that $\Delta t_k=t_{k+1}-t_k<0$, and $v^k$ denotes the approximation of $v$ at $t=t_k$. Particularly, when we set $\theta=\frac12$, the scheme \eqref{dis_sys2v} becomes the famous Crank-Nicolson scheme and is of second-order accuracy; when we set $\theta=1$, the scheme \eqref{dis_sys2v} becomes the backward Euler scheme and is of first-order accuracy.

Given the initial condition of $v$ at $t=T$, we can solve the values of $v$ at the discrete points $(t_k,E_i)$ by using \eqref{dis_sys2v}.

The following theorem declares that the system matrix of system \eqref{dis_sys2v} is an $M$-matrix.
\begin{theorem}
For any given $k=1,2,\cdots,K-1$, if $|\Delta t_k|$ is sufficiently small and $c\geq0$, the system matrix of \eqref{dis_sys2v} is an $M$-matrix.
\end{theorem}

{\bf Proof.} First, we note that $e_{i,j} \leq 0$ for all $i,j=1,2,\cdots,N-1, j\neq i,$ since
\begin{equation}\label{pro}
\frac{b_{i+\frac12}}{e^{\alpha_{i}E_{i+1}}-e^{\alpha_{i}E_{i}}}>0
\end{equation}
for any $i$ and any $b_{i+\frac12}\neq 0$. This is because the function $e^{\alpha E}$ is increasing when $b>0$ and decreasing when $b<0$, where $\alpha=\frac{b}{a}$ and $a=\frac{1}{2}\sigma^2$.
Moreover, \eqref{pro} also holds when $b_{i+\frac{1}{2}}\rightarrow0$. Furthermore, from \eqref{eleft}--\eqref{eright} we know that when $c_{i}\geq0$, for all $i=1,\cdots,N-1$, there holds
\begin{align*}
(e_{i,i})^{k+1}&\geq |(e_{i,i-1})^{k+1}|+|(e_{i,i+1})^{k+1}|+c_{i}^{k+1}l_i.
\end{align*}
Therefore, $D(E,\tau,t^{k+1})$ is a diagonally dominant with respect to its columns. Hence, from the above analysis, we see that for all admissible $i$, $D(E,\tau,t)$ is a diagonally dominant matrix with positive diagonal elements and non-positive off-diagonal elements. This implies that $D(E,\tau,t)$ is an $M$-matrix.

Second, $G^k$ of the system matrix \eqref{dis_sys2v} is a diagonal matrix with positive diagonal entries. In fact, when $|\Delta t_k|$ is sufficiently small, we have
$$
\theta c_{i}l_i+\frac{l_{i}}{-\Delta t_k}>0,
$$
which demonstrates that $\theta D(E,\tau,t) + G$ is an $M$-matrix.\qed

Similarly, we can also discretize the third equation of \eqref{equilibira_system} by using the above method. The parameters in discrete scheme can be modified as follows:

Let $\bar{b}=\tau$, $\bar{\alpha}=\bar{b}/a$ and $\bar{c}=0$. Then, the coefficients are given by
\begin{align*}
\bar{e}_{i,i-1}&=-\bar{b}_{i-\frac12}\frac{e^{\bar{\alpha}_{i-1}E_{i-1}}}{e^{\bar{\alpha}_{i-1}E_{i}}-e^{\bar{\alpha}_{i-1}E_{i-1}}},\\
\bar{e}_{i,i}&=\bar{b}_{i-\frac12}\frac{e^{\bar{\alpha}_{i-1}E_{i}}}{e^{\bar{\alpha}_{i-1}E_{i}}-e^{\bar{\alpha}_{i-1}E_{i-1}}}+\bar{b}_{i+\frac12}\frac{e^{\bar{\alpha}_{i}E_{i}}}{e^{\bar{\alpha}_{i}E_{i+1}}-e^{\bar{\alpha}_{i}E_{i}}}+\bar{c}_il_i,\\
\bar{e}_{i,i+1}&=-\bar{b}_{i+\frac12}\frac{e^{\bar{\alpha}_{i}E_{i+1}}}{e^{\bar{\alpha}_{i}E_{i+1}}-e^{\bar{\alpha}_{i}E_{i}}}.
\end{align*}
These elements can build a matrix $\bar{D}=(\bar{D}_1,\bar{D}_2,\cdots,\bar{D}_{N-1})^{\top}$, where
\begin{align*}
\bar{D}_1&=(\bar{e}_{1,1},\bar{e}_{1,2},0,\cdots,0),\\
\bar{D}_{i}&=(0,\cdots,0,\bar{e}_{i,i-1},\bar{e}_{i,i},\bar{e}_{i,i+1},0,\cdots,0), ~~ i=2,3,\cdots,N-2,\\
\bar{D}_{N-1}&=(0,\cdots,0,\bar{e}_{N-1,N-1},\bar{e}_{N-1,N}).
\end{align*}

Consequently, the numerical discrete scheme for the Kolmogorov equation reads as
\begin{equation}\label{dis_sys2m}
(\theta \bar{D}(E,\tau,t^{k}) + G^{k})m^{k}=(G^k-(1-\theta)\bar{D}(E,\tau,t^{k+1}))m^{k+1}.
\end{equation}
\subsection{The algorithm for \eqref{equilibira_system}}
In the above discussions, we have assumed that the control variable $\tau$ is known, and the density $m$ and the adjoint state $v$ can be solved sequentially. However, we can see from the second equation of \eqref{equilibira_system} that $\tau$ is coupled with $v$. For this reason, we take an iterative method to solve the three unknown functions $\tau$, $m,$ and $v$. The algorithm is presented as follows.
\begin{itemize}
\item {\bf Algorithm}
\end{itemize}

{\bf Step 1}: Give the initial guess of $\tau$, and set it as $\tau^0$. Set a tolerance threshold $Tol>0$ and $n=0$;

{\bf Step 2}: Solve equation \eqref{dis_sys2m} to obtain $m^n$;

{\bf Step 3}: Solve equation \eqref{dis_sys2v} to obtain $v^n$;

{\bf Step 4}: Use $v^n$ to compute $\tau^{n+1}$. Compute
$$
\epsilon^n=\max_{i,k}\|(\tau_i^k)^n-(\tau_i^k)^{n+1}\|;
$$

{\bf Step 5}: If $\epsilon^n\leq Tol$, let $\tau=\tau^{n+1}$, $m=m^n$, $v=v^n$, and stop. Otherwise, let $n=n+1$, and go to Step 2.
\section{Numerical results}\label{results}
So far, we have been able to show the results of our differential game model numerically. We use the following parameter values to solve \eqref{equilibira_system}.

Parameters: $T=1$, $E_{\min}=1$, $E_{\max}=5$, $c_1=10$, $c_2=0.1$, $\sigma=0.3$, $S=0.2$, $E_{0}=1$, $r=0.1$.

\subsection{The efficiency of the numerical method}
\begin{figure}[H]
\centerline{\includegraphics[scale=0.8,trim=170 270 170 280]{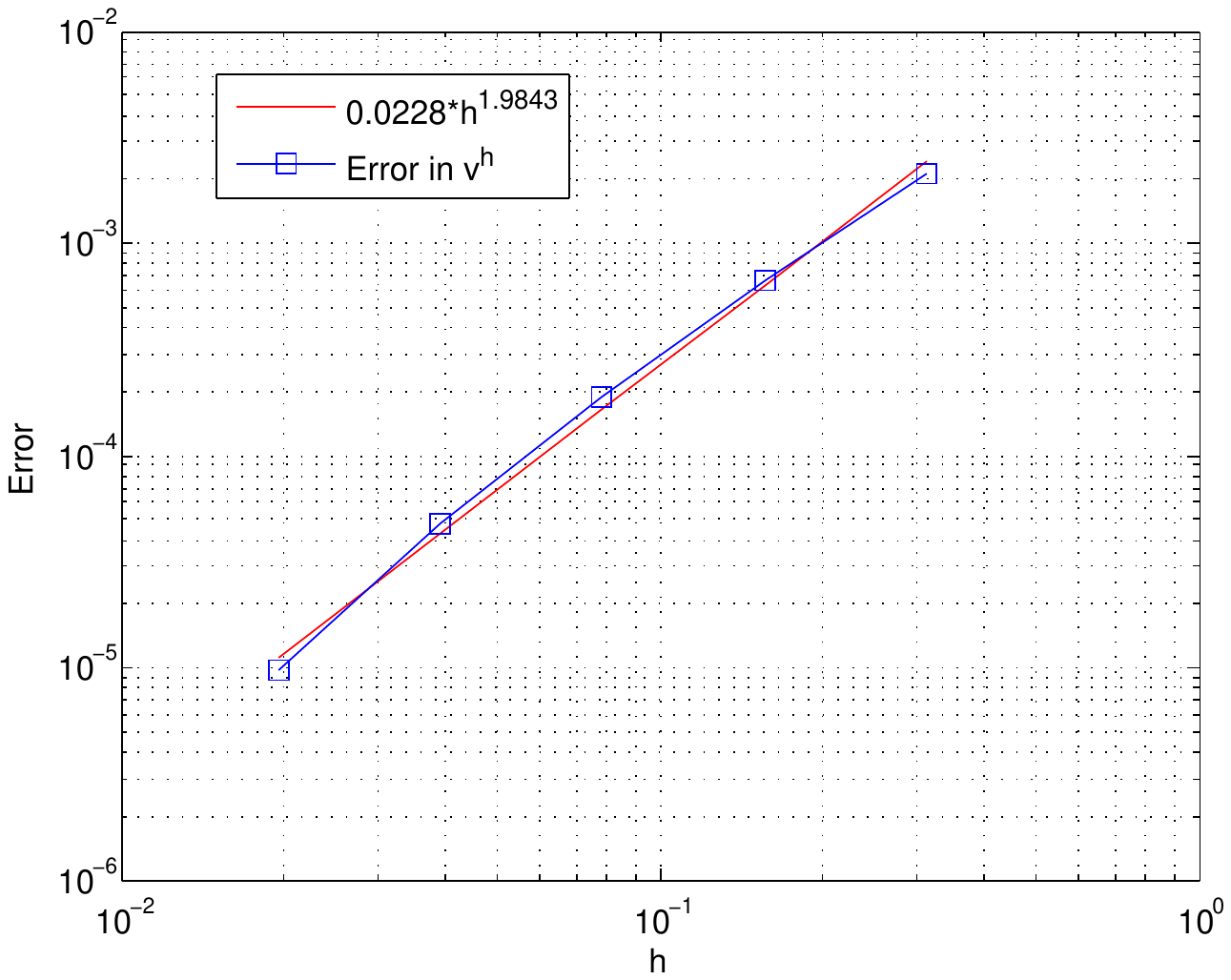} }
\caption{Computed errors in the $L^{\infty}$-norm at $t=0$.}
\label{error}
\end{figure}

First of all, we consider the convergence rate of our discretization method to show its accuracy and efficiency. Owing to the limitation of space, we only test the adjoint state $v$. Additionally, since the closed-form solution of \eqref{equilibira_system} can not be found, we regard the solution obtained by choosing $N=K=512$ in both space and time, respectively, as the ``exact'' solution $v$. We compute the errors in the discrete $L^{\infty}$-norm at the computational final time step $t=0$ on a sequence of meshes with $N=K=2^{n}$ for a positive integer $n$ from $n=4$ to a maximum $n=8$. The discrete $L^{\infty}$-norm is defined as:
$$
\|v^{h}(E,0)-v(E,0)\|_{\infty}=\max \limits_{1<i<N}|v^{h}(E_i,0)-v(E_i,0)|,
$$
where $v^{h}$ denotes the numerical solution. The $\log$-$\log$ plots of the computed maximum errors, along with the linear fitting, are depicted in Figure \ref{error}. From the figure we can see that the rate of convergence of $v^h$ in the discrete $L^{\infty}$ norm is of the order $\mathcal{O}(h^{1.9843})$, where $h=\max \limits_{1<i<N}(h_{i})$. Additionally, it demonstrates numerically that our numerical method for \eqref{equilibira_system} governing the mean field game is useful and efficient. Some theoretical analysis about the stability and the convergence will be discussed in the future works.
\subsection{The solution of the model}
In this subsection, two examples will be addressed to show the equilibrium states under different conditions.

In the first one, the initial density of agents $m_0$ follows a normal distribution, whose mean and variance are 3 and 0.35, respectively. In this case, most of the agents initially choose a medium emission level, and the number of agents whose emission level are maximum or minimum tends to zero.

\begin{figure}[H]
\centerline{\includegraphics[scale=1,trim=170 270 170 270]{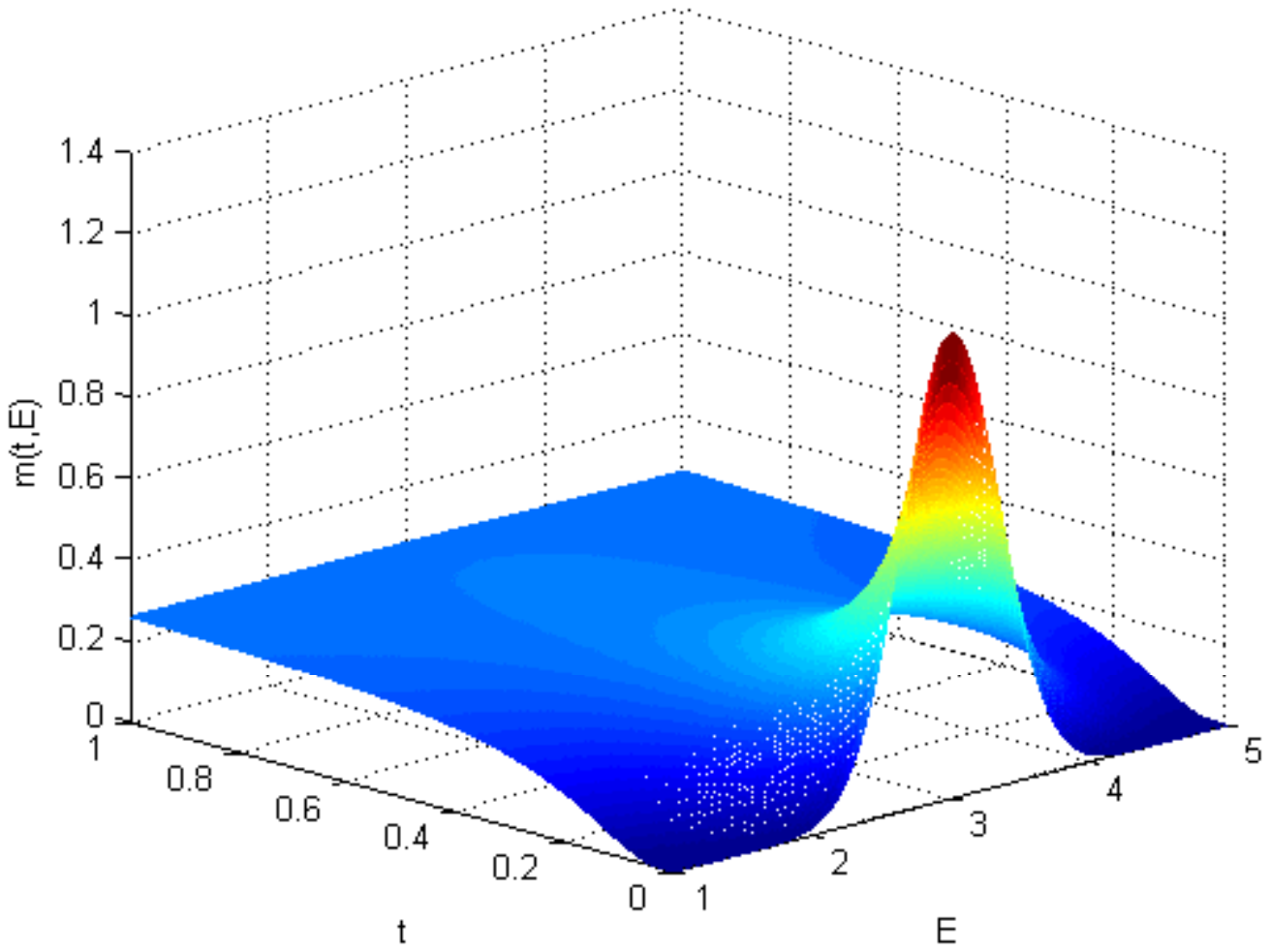} }
\caption{Evolution of $m$.}
\label{normal_m}
\end{figure}

Figure \ref{normal_m} shows the evolution of density for this example. Clearly, we can see from figure \ref{normal_m} that as time goes on, the density of agents averagely disperses on the emission interval $[E_{\min},E_{\max}]$ instead of concentrating at the medium emission level as the initial state did. As mentioned above, the influence of mean field composed of all producers on each agent can be explained by the economic concept ``negative externality''. That is to say, the revenues of each agent should decrease if there are a large number of agents at the same emission level. To maximize own revenues, each agent tends to choose a different emission level from others'.

\begin{figure}
\centerline{\includegraphics[scale=0.95,trim=170 270 170 280]{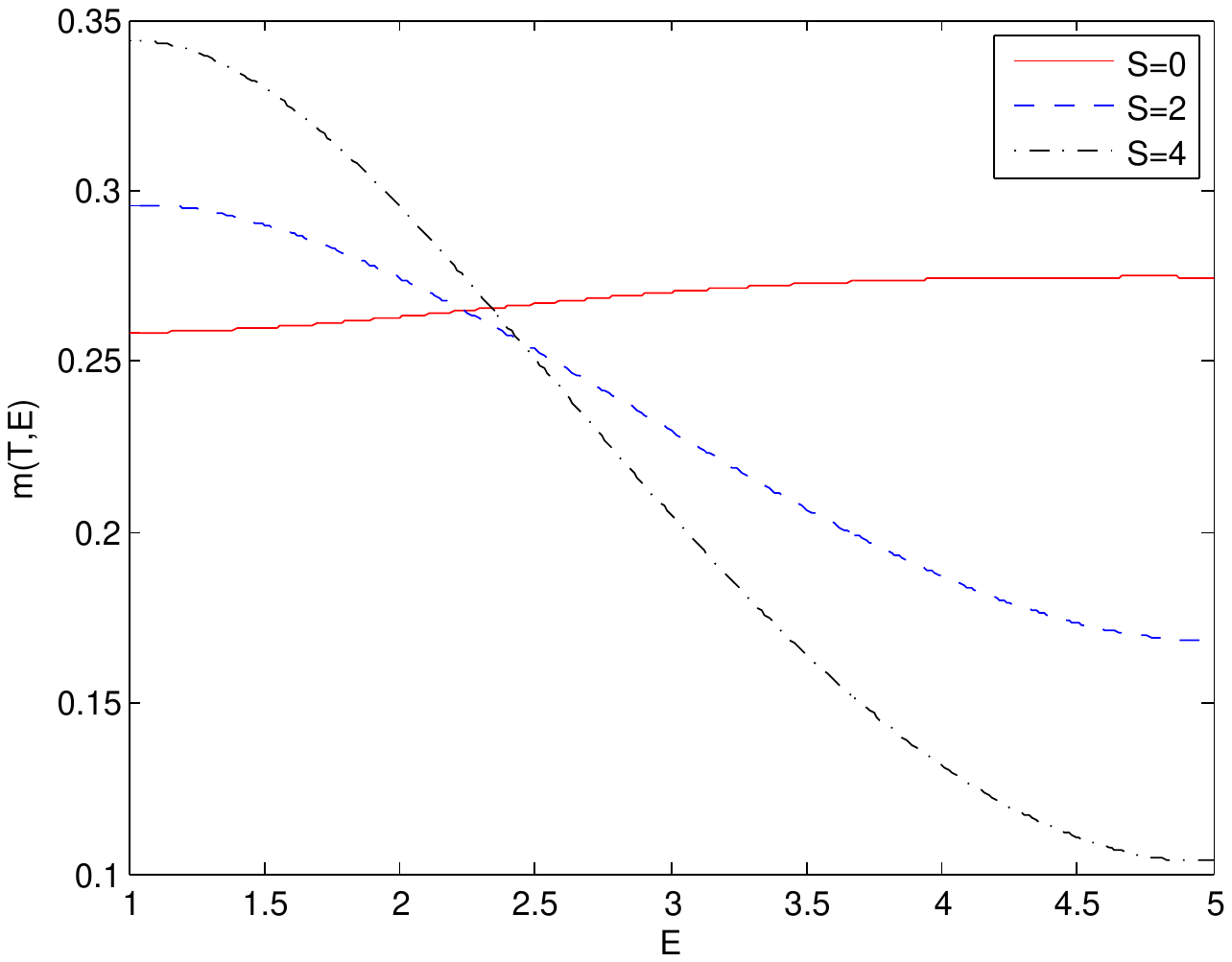} }
\caption{Effect of $S$ on $m(T,E)$.}
\label{normal_S}
\end{figure}

The effect of emission permits price on the equilibrium state is presented in Figure \ref{normal_S}. In the figure, the horizontal axis is the emission level $E$, and the vertical axis is the density of agents at the final time $t=T$, namely $m(T,E)$. We plot three results in the cases of $S=0$, $S=2,$ and $S=4$, respectively. Obviously, with the increasing of emission permits price, the density in lower emission level increases, and the one in higher emission level decreases. In other words, more population would like to choose a lower rather than a higher emission level when the emission permits are expensive.

\begin{figure}
\centerline{\includegraphics[scale=0.95,trim=170 270 170 280]{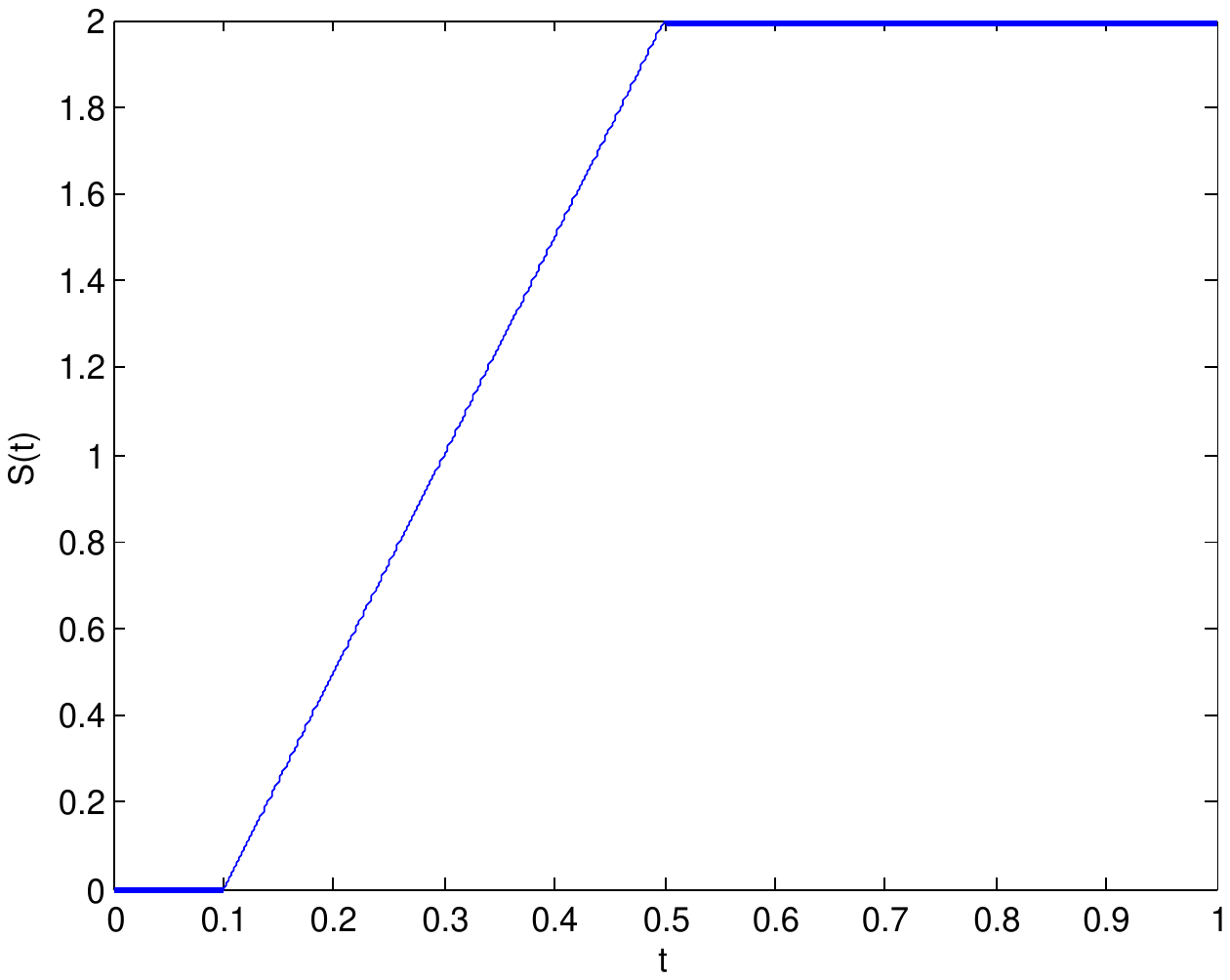} }
\caption{Evolution of emission permits price $S(t)$.}
\label{small_St}
\end{figure}

In the second example, we consider a situation in which most of the agents initially stay together at a lower emission level. The initial density of agents is piecewise linear on $[1,2)$ and $[2,5]$. In addition, the emission permits price becomes time-dependent instead of constant. It is zero at the beginning $[0,0.1)$, then increases from zero to maximum level $S_{\max}=2$ at $[0.1,0.5),$ and keep this level to the end of the game, which is shown in Figure \ref{small_St}.

\begin{figure}
\centerline{\includegraphics[scale=0.95,trim=170 270 170 280]{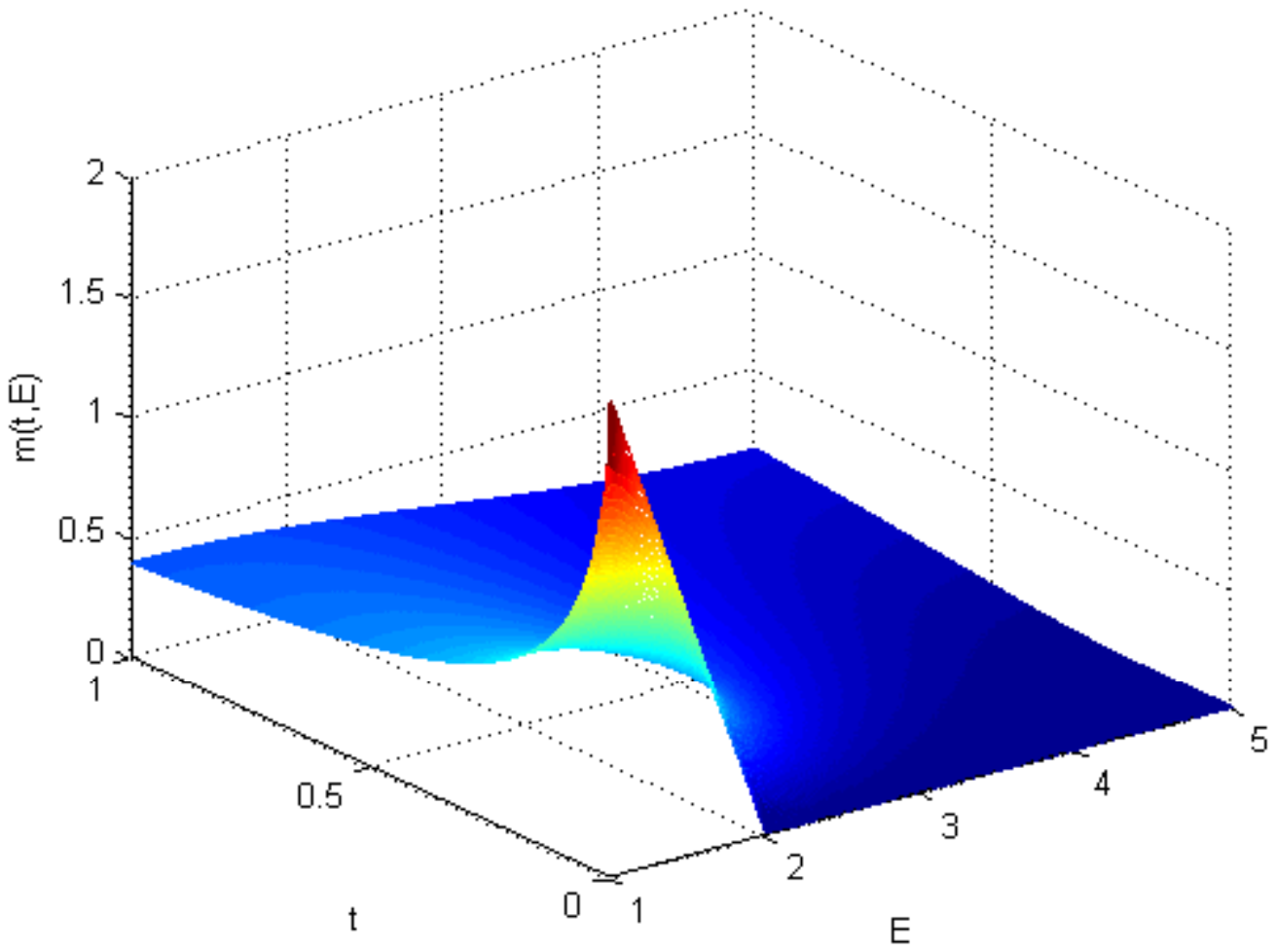} }
\caption{Evolution of $m$.}
\label{small_m}
\end{figure}

The equilibrium density of this example is presented in Figure \ref{small_m}. Similar to the first example, the original obvious aggregation disappears after a period of time. In addition, although the initial condition $m_0$ is non-differentiable at $E=2$, there is no irregular disturbance in the neighbourhood of this point. This can reflect the stability of our algorithm more or less.

\begin{figure}
\centerline{\includegraphics[scale=1,trim=170 270 170 280]{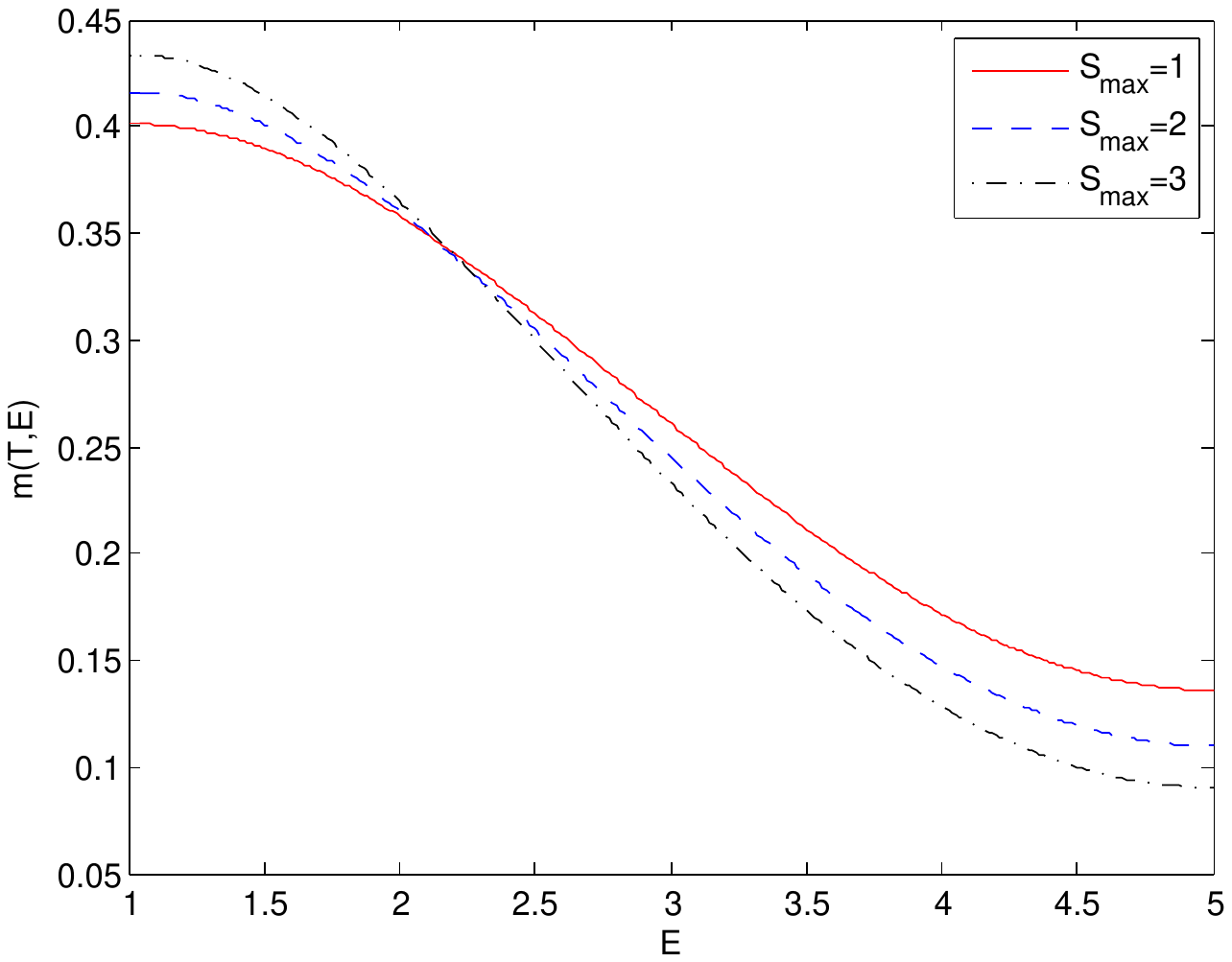} }
\caption{Effect of $S_{\max}$ on $m(T,E)$.}
\label{small_S}
\end{figure}

Furthermore, Figure \ref{small_S} shows the effect of the maximum emission permits price $S_{\max}$ on the equilibrium. Here three cases, namely $S_{\max}=1$, $S_{\max}=2,$ and $S_{\max}=3$, are considered. The similar conclusion to the first example can be obtained. That is, the number of agents in the lower emission level should be increasing as the maximum emission permits price level rises.

\section{Concluding remarks}
We present a mean field game model to study the producers' behaviors in an emission permits trading scheme. In our model, there are a continuum of producers, and each producer is homogeneous. The equilibrium of this game can be represented by a system containing a forward Kolmogorov equation coupled with a backward HJB equation. We then propose a so-called fitted finite volume method to discretize the resulted partial differential equations and the corresponding system matrix is proved to be an $M$-matrix. The efficiency and the usefulness of this method are illustrated by the numerical experiments. Our results show that the convergence rate of our method is nearly $\mathcal{O}(h^2)$. Besides, due to the externality, each agent tends to choose a different emission level from the others'. Finally, we find that the number of agents in lower emission level should be increase when the emission permits are expensive.

We anticipate that our methodology from the perspective of partial differential equations combined with numerical methods can make a few contributions to the solving of complex problems arising from the interdisciplinary fields.

\bibliographystyle{nonumber}

\end{document}